# Dynamic Stark Effect in

# Strongly Coupled Microcavity Exciton-Polaritons

Alex Hayat[1*], Christoph Lange[1], Lee A. Rozema[1], Ardavan Darabi[1], Henry M. van Driel[1], Aephraim M. Steinberg[1], Bryan Nelsen[2], David W. Snoke[2], Loren N. Pfeiffer[3] and Kenneth W. West[3]

[1]*Department of Physics, Centre for Quantum Information and Quantum Control, and Institute for Optical Sciences, University of Toronto, Toronto, Ontario M5S 1A7, Canada*
[2]*Department of Physics and Astronomy, University of Pittsburgh, Pittsburgh, Pennsylvania 15260, USA*
[3]*Department of Electrical Engineering, Princeton University, Princeton, New Jersey 08544, USA*
*email: ahayat@physics.utoronto.ca*

We present experimental observations of a non-resonant dynamic Stark shift in strongly coupled microcavity quantum well exciton-polaritons – a system which provides a rich variety of solid-state collective phenomena. The Stark effect is demonstrated in a GaAs/AlGaAs system at 10 K by femtosecond pump-probe measurements, with the blue shift approaching the meV scale for a pump fluence of 2 mJcm$^{-2}$ and 50 meV red detuning, in good agreement with theory. The energy level structure of the strongly coupled polariton Rabi-doublet remains unaffected by the blue shift. The demonstrated effect should allow generation of ultrafast density-independent potentials and imprinting well-defined phase profiles on polariton condensates, providing a powerful tool for manipulation of these condensates, similar to dipole potentials in cold atom systems.

Strong coupling of matter to a vacuum mode has been a growing field of study in atomic physics since the demonstrations of vacuum Rabi splitting [1], which manifests the new dressed eigenstates of the system. In semiconductors, these dressed states correspond to new quasi-particles when quantum well (QW) excitons are strongly coupled to a microcavity mode [2]. The exciton-polaritons which emerge from this strong coupling have a unique combination of extremely low effective mass and relatively strong polariton-polariton interactions which enable observations of a wide range of physical phenomena including strong parametric scattering [3], and high-temperature Bose-Einstein condensation (BEC) [4,5,6,7] which exhibits Bogoliubov excitations [8] and superfluid behavior [9,10,11].

Matter-wave interferometry in cold-atom BEC has been demonstrated by using the dynamic Stark shift to generate controllable potential landscapes [12], and to imprint spatial phase profiles on the condensate. The imprinted phase profiles have been also shown to result in the generation of solitons [13] and vortices [14]. Optical manipulation of exciton-polariton BEC was recently achieved by potentials created by repulsive interaction between excitons, showing long-range order [15,16] as well as generation and control of quantum-fluid vortices [17]. The switching time of these potentials based on carrier generation, however, is limited by the carrier dynamics, preventing fast manipulation and characterization of the condensate. Moreover, such density-dependent potentials make the extraction of the condensate properties from interference experiments much more difficult. Previously, only predetermined static density-independent potentials, which are unsuitable for phase imprinting and condensate manipulation, have been demonstrated [18].

Generating dynamic Stark-based potentials for exciton-polariton BEC can provide a fast-switching and density-independent technique for manipulating the condensate and inducing well-defined phase profiles by time-dependent potentials. Resonant high-intensity excitation of strongly coupled exciton-polaritons was shown to modify the energy level structure significantly from a Rabi doublet to a Mollow triplet [19]. However, a non-resonant dynamic Stark shift of a strongly coupled light-matter interaction system has not been observed previously, and an experimental demonstration of this effect paves the way towards manipulation of the vacuum-dressed states through an additional external-field dressing.

Here we present observations of the non-resonant dynamic Stark shift in a strongly coupled light-matter system. The effect is demonstrated on microcavity exciton-polaritons by fs pump-probe spectroscopy showing a significant blue shift of both the lower polariton (LP) and the upper polariton (UP) energies. Our experiments yield meV-scale blue shifts, similar to potentials currently used in exciton-polariton BEC experiments [15-18], at readily available pump fluences of approximately 2mJcm$^{-2}$ with ~50 meV red detuning. The fs Stark shift in our experiments corresponds to a radian scale phase imprinting, while longer pulses at similar intensities can result in larger phase. The polariton energy level structure remains intact in our experiments with the Rabi splitting not affected by the blue Stark shift. The intensity dependence of the shift is in good agreement with our theoretical modeling.

The exciton-polariton dressed state dispersion in the strongly coupled microcavity is determined by diagonalizing the Hamiltonian including the interaction resulting in [20]

$$E_{LP,UP}(k) = \frac{1}{2}\left( \begin{array}{c} E_X(k) + E_C(k) + i(\gamma_X + \gamma_C) \pm \\ \pm\sqrt{(\hbar\Omega_0)^2 + \left[E_X(k) - E_C(k) + i(\gamma_C - \gamma_X)\right]^2} \end{array} \right) \quad (1)$$

where $\hbar k$ is the in-plane momentum, $E_X(k)$ and $E_C(k)$ are the uncoupled exciton and cavity dispersion relations, respectively, $\gamma_X$ and $\gamma_C$ are determined by the uncoupled exciton and cavity photon decay rates and the inhomogeneous broadening, and $\Omega_0$ is the Rabi frequency of the exciton coupled to the cavity mode. These dressed dispersion relations of the strongly coupled system define the LP and the UP, whereas the pump-induced dynamic Stark blue shift yields an additional dressing (Fig. 1 a).

The dynamic Stark shift of bare QW exciton energy levels on a fs time scale, with no coupling to an optical cavity, has been shown previously with strong dependence on the red detuning of the pump with the residual absorption of the pump spectrum limiting the minimal detuning [21,22]. However, to the best of our knowledge, dynamic Stark shift of any strongly-coupled system has not been observed before, taking our observations to a new regime of light-matter interaction. Furthermore, our theoretical modeling shows that for sufficiently strong intensities, the shift of the strongly coupled polariton, as a single particle, is different from the shift of just the exciton component of the polaritons.

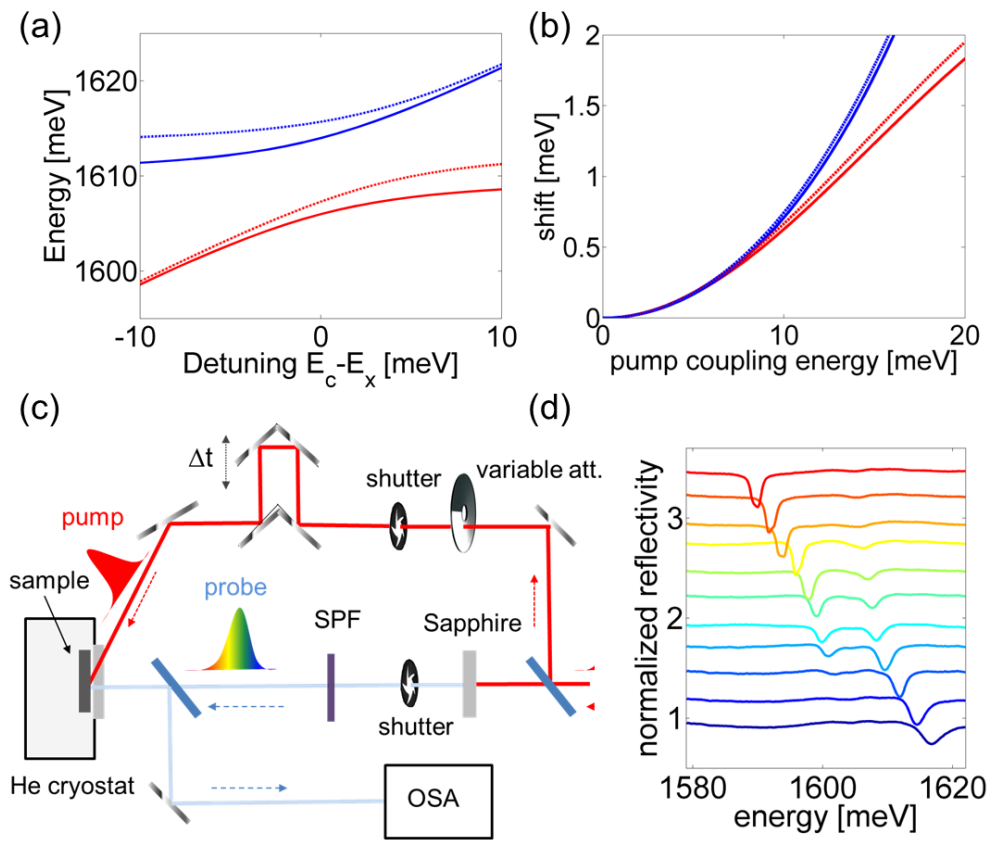

**Figure 1.** (color online) (a) Calculated polariton levels using full Hamiltonian diagonalization (Eq. 2) with no pump (solid lines) and pump coupling energy of 15 meV (dashed lines) for the UP (blue) and the LP (red). (b) Calculated dynamic Stark shift vs. pump coupling energy using full Hamiltonian diagonalization (blue) and exciton Stark shift approximation (red) for the UP (solid lines) and the LP (dashed lines). (c) Schematic diagram of the experimental setup. (d) Measured normalized reflectivity spectra at normal incidence for different cavity-exciton detuning. The curves are offset for convenience.

The effect of the dynamic Stark shift of vacuum-dressed exciton-polaritons by an additional dressing by the pump field is described fully by diagonalizing the Hamiltonian that includes both the vacuum interaction and the non-resonant pump field (neglecting the decay rates),

$$H = \begin{pmatrix} E_X(k) & \hbar\Omega_0 & \hbar\Omega_p \\ \hbar\Omega_0 & E_C(k) & 0 \\ \hbar\Omega_p & 0 & \hbar\omega_p \end{pmatrix} \qquad (2)$$

where $\omega_p$ is the radial frequency of the pump laser, and $\Omega_p = d|\mathscr{E}|/\hbar$ is the corresponding Rabi frequency, with the dipole matrix element $d$ and $\mathscr{E}$ the pump electric field. Diagonalization of the Hamiltonian (Eq. 2) results in blue shifted UP and LP levels (Fig.1 b). The exciton-polariton is affected by the Stark shift as a single particle, so that the Rabi splitting remains intact. Treatment of the polaritons perturbatively by calculating UP and LP levels with only the exciton level shifted by the dynamic Stark effect yields slightly different level shifts. This difference becomes more significant for higher pump intensities where the pump-induced Rabi frequency becomes comparable to the vacuum splitting (Fig.1 b). For weaker pump fields, the polariton Stark shift can be approximated by the change in polariton energies due to the Stark shift of the excitons only, and for our experimental conditions the error due to this approximation is < 6% - comparable to experimental error. Dressing uncoupled excitons due to Stark shift results in a pump-induced level splitting given by [23]

$$\Delta E_X = \sqrt{\left(E_X - \hbar\omega_p\right)^2 + \left(\hbar\Omega_p\right)^2} \qquad (3)$$

The resulting polariton dispersion relations including the Stark dressing can therefore be calculated:

$$E_{LP,UP}(k) = \frac{1}{2}\left[ \begin{array}{l} \sqrt{\left(E_X(k) - \hbar\omega_p\right)^2 + \left(\hbar\Omega_p\right)^2} + \\ + \hbar\omega_p + E_C(k) + i\left(\gamma_X + \gamma_C\right) \pm \\ \pm \left( \left(\hbar\Omega_0\right)^2 + \left[ \begin{array}{l} \sqrt{\left(E_X - \hbar\omega_p\right)^2 + \left(\hbar\Omega_p\right)^2} + \\ + \hbar\omega_p - E_C(k) + i\left(\gamma_C - \gamma_X\right) \end{array} \right]^2 \right)^{\frac{1}{2}} \end{array} \right]^{\frac{1}{2}}. \qquad (4)$$

In our experiments, the ultrafast dynamic Stark shift in a strongly coupled exciton-polariton sample was measured by pump-probe differential reflection spectroscopy in a liquid He flow microscopy cryostat at a temperature of 10 K (Fig. 1 c). The output of a regenerative Ti:sapphire amplifier with a repetition rate of 250 kHz was used to induce the shift with 800 nm, 250 fs pulses, and to generate white-light supercontinuum pulses in a sapphire crystal for the probe. Pump and probe were cross-linearly polarized to suppress scattered pump light in the detector.

The strongly coupled exciton-polariton structure was grown by molecular beam epitaxy, and consisted of a $\lambda/2$ AlAs layer embedded between two $Ga_{0.8}Al_{0.2}/AlAs$ distributed Bragg reflectors (DBR). The DBRs consist of 20 layer pairs for the bottom reflector and 16 pairs for the top reflector, resulting in a high-quality $3\lambda/2$ microcavity. Our structure is similar to the one used in [4], with the difference that a single (instead of four) 6.5 nm GaAs QW is placed at each of the 3 antinodes of the microcavity. Therefore the two AlAs barriers of the central QW are parts of the $\lambda/2$ layer – 63 nm thick, while the other two QWs have one 10nm and one 59 nm AlAs barrier. The thickness of the layers was tapered across the sample to allow tuning of the cavity resonance by probing at different locations on the sample. At normal incidence, the pump wavelength of 800 nm is strongly reflected by the DBR. In order to maximize the coupling of the pump pulse into the microcavity, the pump beam was incident at an angle of approximately $70^{\circ}$. The strongly coupled region of the sample was identified by reflection spectroscopy with the LP dip at 775 nm and the UP at 771 nm (Fig. 1 d). The thickest GaAs layers in the structure (besides the substrate) are the 6.5 nm QWs with absorption at a higher energy than the pump. The thick layers in the structure are $Al_{0.2}Ga_{0.8}As$ which have a bandgap of ~1.8 eV at 10 K and do not absorb the pump, confirmed by the lack of a direct-absorption-induced ns time-scale luminescence signal. A time-independent shift due to heating of the substrate, more than an order of magnitude smaller than the dynamic Stark shift, was removed from the measurements by subtracting the constant background.

Differential measurements of the dynamic Stark shift were performed, where the pump-induced change in reflectivity was measured. For each measurement, four spectra were recorded to reduce background and photoluminescence effects: both pump and probe were selectively blocked or unblocked; the four possible combinations yield the background signal $I_{BG}$ (both off), the white-light reference $I_{PR}$ (pump off), the photoluminescence $I_P$ (probe off) and the actual Stark shift measurement $I_{PPR}$ with both beams switched on. The normalized differential reflection spectrum was extracted from the measurements through

$$\frac{\Delta R}{R} = \frac{I_{PPR} - I_P}{I_{PR} - I_{BG}} - 1 \tag{5}$$

.

The two polariton resonances appear as separate reflection features. Differential absorption spectroscopy renders the difference in reflectivity of the initial and the modified reflectivity features. The shift caused by the dynamic Stark effect thus leads to a decrease of relative reflectivity above the polariton resonance, and an increase below the resonance. Since two resonances are involved, two such dispersive features arise. For spectral shifts smaller than the linewidth of UP and LP, the shift appears as a dispersive feature in the normalized differential reflection $\Delta R/R$, and both the amplitude and width of the $\Delta R/R$ features are functions of the amplitude and width of UP and LP reflection dips. Therefore the magnitude of the dynamic Stark shift determines the amplitude of the differential reflection feature.

Time-dependent normalized differential spectrum measurements obtained from the pump-probe experiments clearly demonstrate the dynamic Stark blue shift of both LP and UP lines (Fig. 2 b-e), showing good agreement with the calculated spectra (Fig. 2 a). The maximal dynamic Stark shift is attained at approximately one FWHM of the probe pulse before zero time delay, when most of the probe pulse has been absorbed. The probe-induced polarization has not decohered notably at this point and attains its maximal value. Accordingly, the pump-induced Stark shift has the largest effect here. Our quantitative evaluation of the Stark shift is performed at this delay time. At positive delay times, zero signal is obtained, confirming that the pump pulse does not generate any carriers which would lead to a long-lived change in reflectivity. Before time zero, where the wide-band probe pulse arrives before the pump, coherent oscillations of the probe polarization are observed [24,25]. Also known as perturbed free induction decay, these oscillations result from a finite decay time of the coherent polarization induced at electronic resonances by the probe. The coherent polarization is perturbed by the pump subsequently incident within the coherence time of the polarization [26].

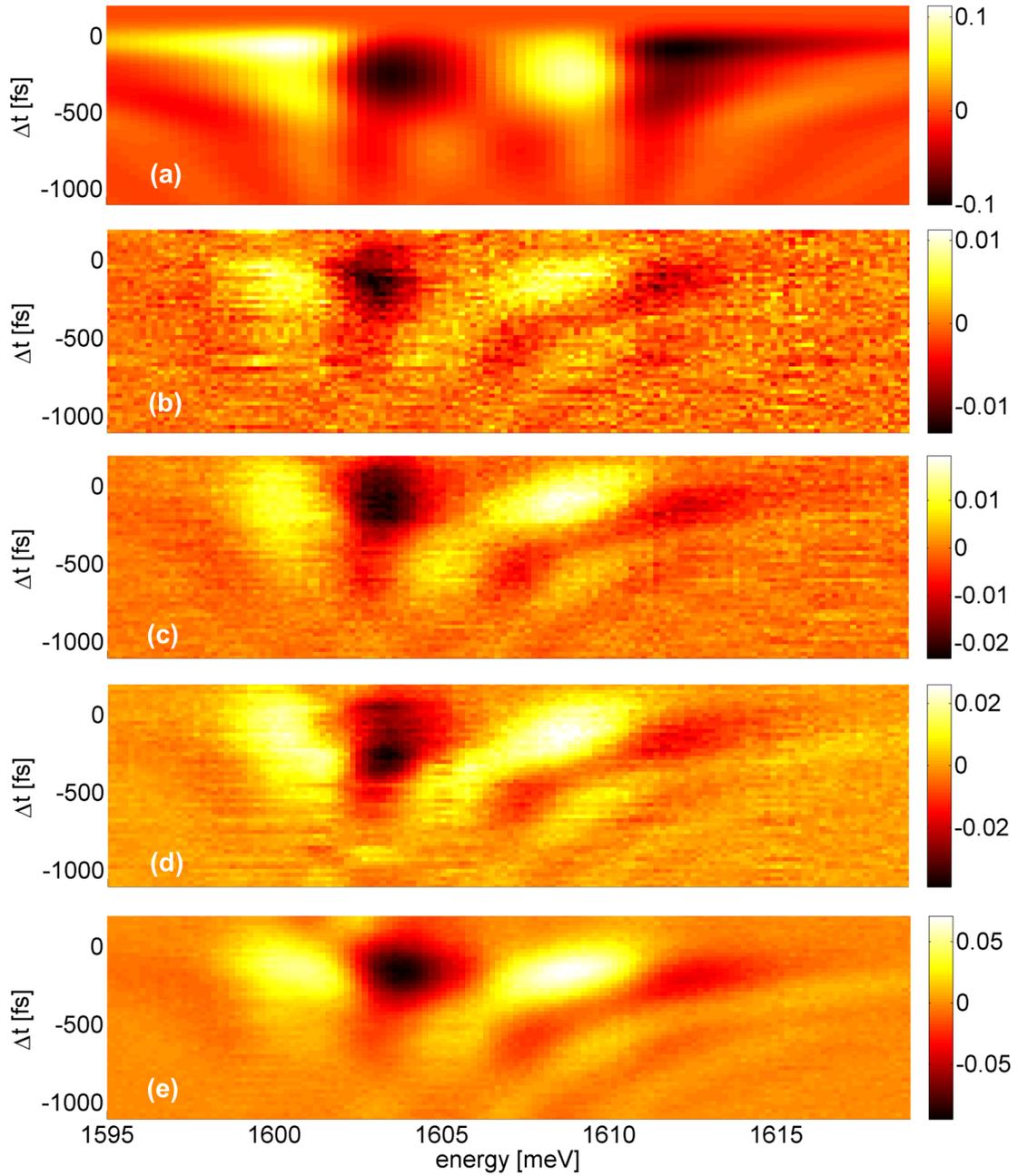

**Figure 2.** (color online) (a) Calculated normalized differential spectra $\Delta R/R$ for various pump-probe time-delays $\Delta t$, for a pump fluence of 0.2 mJcm$^{-2}$. Measured normalized differential reflection $\Delta R/R$ spectra for various pump-probe time-delays for a pump fluence of (b) 0.2 mJcm$^{-2}$ (c) 0.6 mJcm$^{-2}$ (d) 1.2 mJcm$^{-2}$ (e) 2 mJcm$^{-2}$.

The re-radiation of the polarization thus consists of two components with a relative phase shift and intensity ratio - both depending on the pump-probe delay. Both fields add to render the intensity seen by the time-integrating detector, and manifest as oscillatory spectral features. The spectral period of these oscillations diverges at time zero (Fig. 2 b-e), allowing fs-scale determination of the time zero from an inherent property of the electronic system. Our calculations include this coherent pump probe interaction modeled by semiconductor Bloch equations based on a previously developed theory [25], yielding good agreement with the experiment and validating our quantitative analysis.

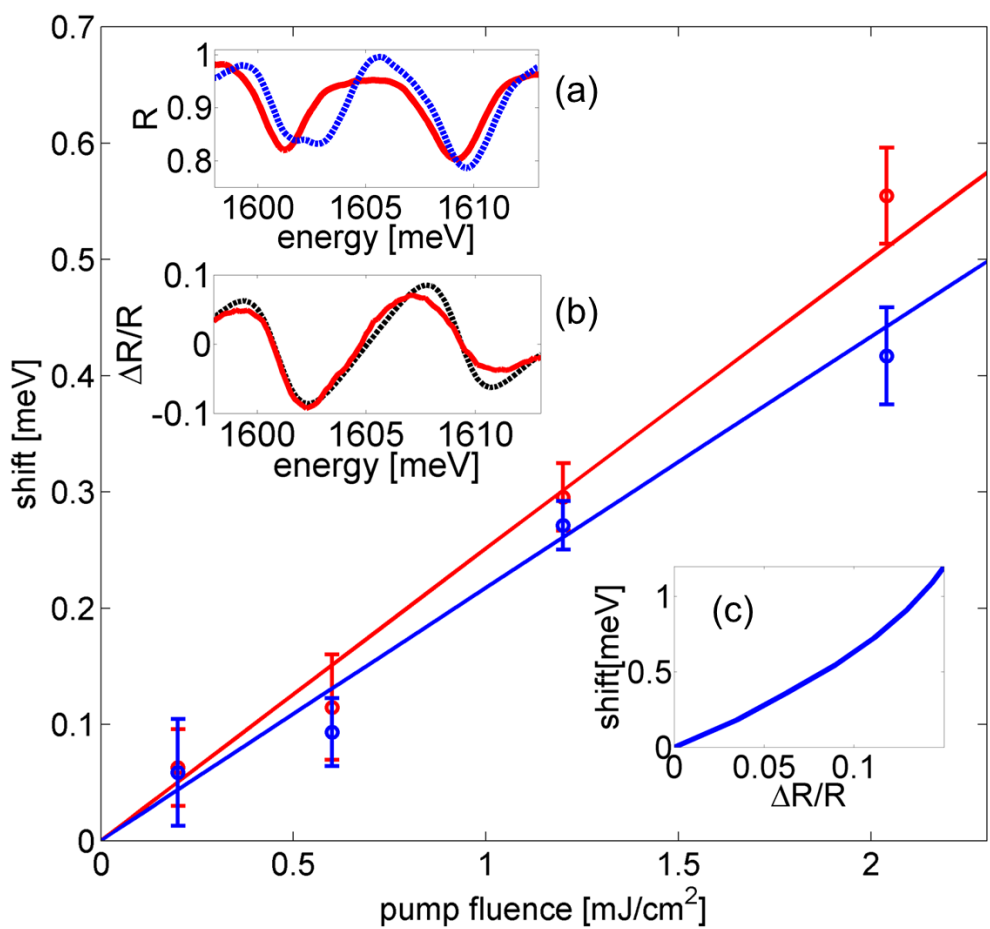

**Figure 3.** (color online) Dynamic Stark blue shift vs. pump fluence. The calculated dependence is given by the solid lines: red for LP and blue for UP, with pump coupling efficiency as a fitting parameter. Inset (a) normalized reflection spectra for the initial (solid red) and the Stark-shifted (dashed blue) polariton levels at a constant delay. Inset (b) is a 2 mJcm$^{-2}$ fluence differential reflection at a constant delay for measured (solid line red) and calculated (dashed black line). Inset (c) shows the shift vs. differential reflection amplitude $\Delta R/R$ as a conversion curve from $\Delta R/R$ (Fig. 2) to the energy shift.

The magnitude of the dynamic Stark shift as extracted from the amplitude of the normalized differential reflection features is shown on Fig. 3. In the intensity range used in our experiments, the shift is less than the Rabi splitting between LP and UP. In contrast to the modified triplet structure appearing in resonant pumping [19], the Rabi splitting is maintained during the fs Stark shift with non-resonant pumping here. Hence, a change in the underlying potential is introduced, which is non-destructive to the exciton-polaritons. The intensity dependence of the measured Stark shift agrees well with our calculations (Fig. 3), and the shift obtained with readily available pulse fluences approaches meV scales. The maximal peak pump power in our experiment was limited by two-photon absorption-induced luminescence from the sample, which starts dominating the spectrum at peak intensities higher than $2 \cdot 10^{10} Wcm^{-2}$.

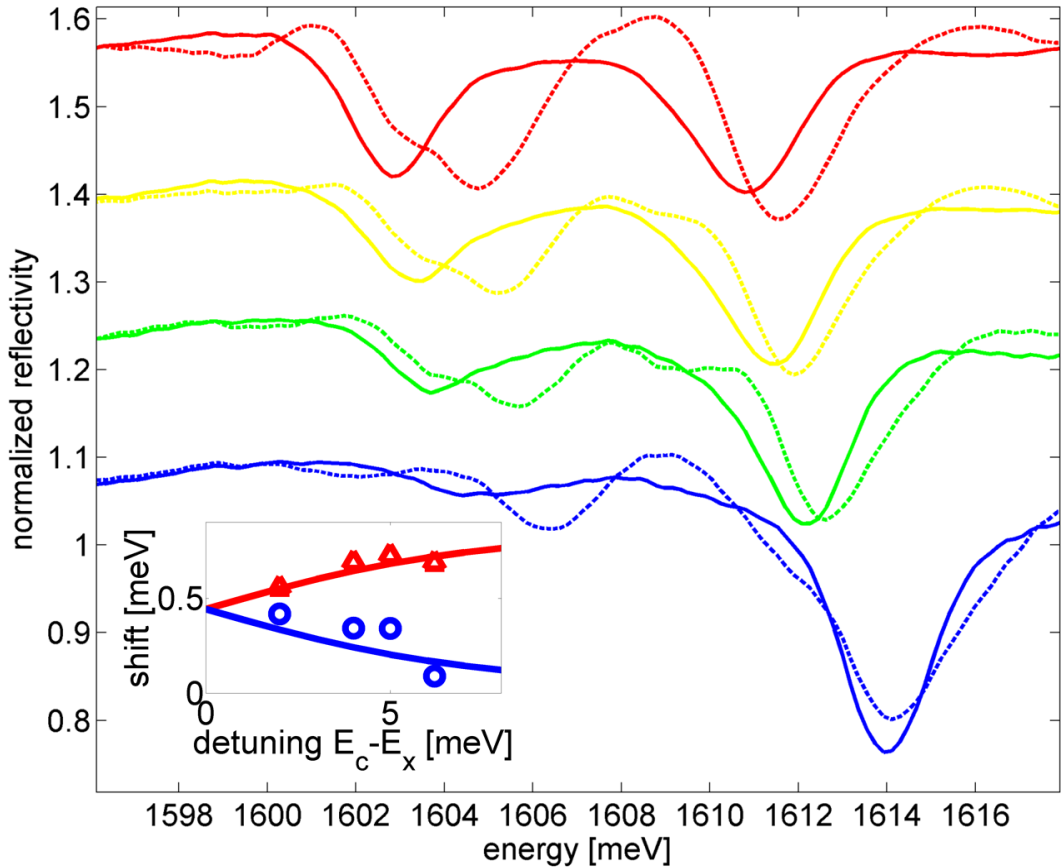

**Figure 4.** Measured normalized reflectivity spectra at normal incidence of the initial (solid) and the Stark-shifted (dashed) polaritons, for different cavity-exciton detuning and constant delay. The curves are offset for convenience. The inset is the Stark shift dependence on cavity-exciton detuning - measured points and calculated curves show the LP (red) and UP (blue).

In order to study the effect of photon-exciton coupling strength on the Stark shift of the polaritons and to rule out a possible Kerr-induced bare cavity resonance shift, we performed several experiments at various cavity-exciton detunings. Such detuning results in two effects: the photon-like polariton feature in reflection is deeper and narrower than the exciton-like one (Fig.1 d), and the exciton-like polariton is affected more by the Stark shift. The difference between the shifts of the UP and LP becomes larger for larger detuning, as well as the difference in the shapes of the features. As expected from the theoretical model (Fig. 1 a), the shift of the exciton-like branch increases with increasing detuning, while the shift of the photon-like branch decreases (Fig. 4). The small additional features appearing in the shifted spectra are due to the coherent oscillations discussed above (Fig. 2). For cavity-exciton detuning as small as 10 meV, the shift of the photon-like branch is strongly reduced (by almost an order of magnitude). This is in good agreement with the calculation of the Stark shift of a polariton with a smaller exciton component (Fig. 4 inset). In contrast, the Kerr effect modifying the cavity index alone should neither depend significantly on detuning, nor, as recently shown for GaAs microcavities [27], manifest as a blue shift as seen here, but as a red shift instead.

Although the demonstrated dynamic Stark shift is smaller than the linewidth of the polaritons, in the condensed state the linewidth becomes much narrower - demonstrated in several recent experiments showing long-range coherence [15,17, 18]. The shifts demonstrated here reach values of 0.5 meV – similar to the potentials used in several previous polariton BEC manipulation experiments [16,17] and larger than demonstrated trapping potentials [18]. Moreover, for phase imprinting, the key quantity is the accumulated phase, which depends on both the energy shift and its duration. In our experiment the 0.5 meV energy shift during 250 fs corresponds to 0.2 radian phase change, such that a phase change of $\pi$ could be obtained with a similar pump during several picoseconds, which is shorter than typical BEC coherence times [16]. Furthermore, with fs pulses, the dynamic Stark shift allows generation of potentials on time scales shorter than polariton lifetimes, which is impossible by alternative methods.

In conclusion, we have demonstrated a non-resonant fs time-scale dynamic Stark shift in a strongly coupled system of exciton-polaritons. The effect is observed by pump-probe differential reflection spectroscopy. We observe a blue shift of the exciton polariton levels approaching the meV energy scale, similar to currently used potentials in polariton BEC and in good agreement with our calculations. The vacuum Rabi splitting of the polaritons remains unaffected, demonstrating the non-invasive nature of this tool for ultrafast control of polariton condensates. The non-resonant dynamic Stark shift presented here is a demonstration of non-resonant dressing of a strongly-coupled vacuum-dressed system, and it thus provides a unique method for ultrafast manipulation of exciton-polariton BEC by density independent potentials enabling various applications in quantum technologies and fundamental studies of spontaneous coherence in matter.

Acknowledgements: We gratefully appreciate financial support from Natural Sciences and Engineering Research Council of Canada, Canadian Institute for Advanced Research and the Alexander von Humboldt Foundation (C.L.). The work at Princeton was partially funded by the Gordon and Betty Moore Foundation as well as the National Science Foundation MRSEC Program through the Princeton Center for Complex Materials (DMR-0819860). The work at the University of Pittsburgh was supported by the National Science Foundation (DMR- 1104383).